\documentclass[10pt, conference]{IEEEtran}
\IEEEoverridecommandlockouts
\usepackage{cite}
\usepackage{amsmath,amssymb,amsfonts}
\usepackage{algorithmic}
\usepackage{graphicx}
\usepackage{textcomp}
\usepackage{xcolor}
\usepackage{booktabs}
\usepackage{multirow}
\usepackage{url}
\usepackage{hyperref}
\usepackage{xcolor}
\usepackage{orcidlink}
\hypersetup{
    colorlinks=true,
    linkcolor=blue,
    citecolor=blue,
    filecolor=magenta,      
    urlcolor=black,
    pdftitle={},
}

\def\BibTeX{{\rm B\kern-.05em{\sc i\kern-.025em b}\kern-.08em
    T\kern-.1667em\lower.7ex\hbox{E}\kern-.125emX}}

\begin{document}


\title{A Modular Event-Driven Software Architecture for Open-Source Industrial IoT Edge Gateways}

\author{
\IEEEauthorblockN{%
  Pei Yu Wong\IEEEauthorrefmark{1},
  Thien Tran\IEEEauthorrefmark{2},
  Hudyjaya Siswoyo Jo\IEEEauthorrefmark{1} and
  Jonathan Kua\IEEEauthorrefmark{2}
}
\IEEEauthorblockA{%
    \IEEEauthorrefmark{1}Swinburne University of Technology Sarawak Campus, Malaysia \\
    \IEEEauthorrefmark{2}School of Information Technology, Deakin University, Australia
    }
\IEEEauthorblockA{
    wpeiyu0116@gmail.com;
    hsiswoyo@swinburne.edu.my;
    \{peter.tran, jonathan.kua\}@deakin.edu.au
    }
}

\maketitle

\begin{abstract}
Integrating legacy industrial machinery into modern cloud infrastructures poses a significant software engineering challenge. Industrial Internet of Things (IIoT) deployments frequently rely on rigid and proprietary edge controllers that require extensive manual configuration. Open-source single-board computers (SBCs) offer a highly adaptable hardware alternative, but they still lack standardized software deployment frameworks capable of bridging localized serial protocols with global cloud environments. In this paper, we present a modular event-driven software architecture designed specifically for open-source IIoT edge gateways. We built a containerized web stack to provide a unified architecture that abstracts the complexities of Modbus-to-MQTT protocol translation, enabling rapid configuration of telemetry streams to cloud platforms such as Amazon Web Services (AWS) IoT. Furthermore, we designed a localized event engine that processes conditional logic directly at the edge layer, which significantly reducing network latency. The proposed architecture is preliminarily deployed and validated on an industrial-grade Raspberry Pi Compute Module 4 (CM4) to investigate its robustness, scalability and user-friendliness. Experimental results demonstrate a substantial reduction in system integration time and high operational stability, thus establishing a robust software foundation for real-time edge-based industrial automation. This paper contributes to extending the lifespan of legacy machinery by enabling seamless integration with modern industrial infrastructure.
\end{abstract}

\begin{IEEEkeywords}
Open-Source Edge Gateway Configurator, Client-Side Software/Application Architecture, Industrial IoT
\end{IEEEkeywords}

\section{Introduction}
Modernizing Industrial Cyber-Physical Systems (ICPS) through Industrial Internet of Things (IIoT) technologies requires efficient data acquisition across heterogeneous network layers~\cite{wu2018iot, liu2019edge}. Recent paradigm shifts in ICPS towards Cloud-Fog Automaton (CFA) has motivated the developments of open-source industrial software to facilitate the seamless data communication across cloud/fog/edge layers~\cite{cfa_tii_24, jin_cfa_jsac25}. The transition toward smart manufacturing introduces a critical bottleneck at the edge layer: Bridging legacy field devices with cloud-native applications~\cite{wang2024sdn}. Traditional field devices communicate via localized and deterministic serial protocols such as Modbus RTU, which requires seamless connectivity to collect, process and analyze data~\cite{schlemitz2024sm, john2020enabling}. On the other hand, industrial analytical platforms require asynchronous protocols such as Message Queuing Telemetry Transport (MQTT) for large-scale data ingestion~\cite{iotvr_jamt_25, coltran_jii_25, ladegourdie2022performance}.

Proprietary Programmable Logic Controllers (PLCs) and edge gateways currently dominate this integration domain in IIoT. These closed-ecosystem solutions provide high reliability but incur high financial costs, limited adaptability and architectural rigidity for Small and Medium-Sized Businesses (SMBs), where flexibility and budget friendliness are prioritized~\cite{chen2022design, jan2022real}. Engineers face substantial barriers when attempting to implement custom algorithms or integrate third-party sensors into these closed environments. Practically, open-source single-board computers (SBCs) (e.g., Raspberry Pi, Beagle Board) present alternatives, offering extensive input/output flexibility and computational power at a fraction of the cost~\cite{johnston2016applicability} while being capable of running full operating systems~\cite{isikdag2015internet}. 

However, deploying SBCs in industrial settings introduces complex software engineering challenges. Technicians traditionally write extensive custom code to initialize network interfaces, parse serial bit-streams, and manage secure cloud data handshakes. This fragmented approach increases deployment time and introduces a high margin for configuration errors. The development of IIoT gateways remains a pivotal application area for enabling flexible cloud connectivity. Hence, SBCs have increasingly focused on replacing costly proprietary hardware with economical open-source alternatives.

Recent studies highlight the efficacy of utilizing development boards for industrial data acquisition. Chen and Hsu~\cite{chen2022design} designed an economical IoT gateway utilizing the LinkIt\texttrademark\ Smart 7688. Their software system successfully transmitted sensor monitoring data to a cloud server via a web interface, demonstrating the feasibility of using Modbus TCP/IP and MQTT on low-cost hardware. Similarly, Sun et al. \cite{sun2019design} presented a Modbus-to-MQTT gateway using a Raspberry Pi coupled with an RS-485 add-on board. They utilized a smart agent to convert field-level data into the MQTT protocol, enabling communication with Alibaba Cloud and AWS. 

\begin{figure*}[htbp]
\centerline{\includegraphics[width=0.9\textwidth]{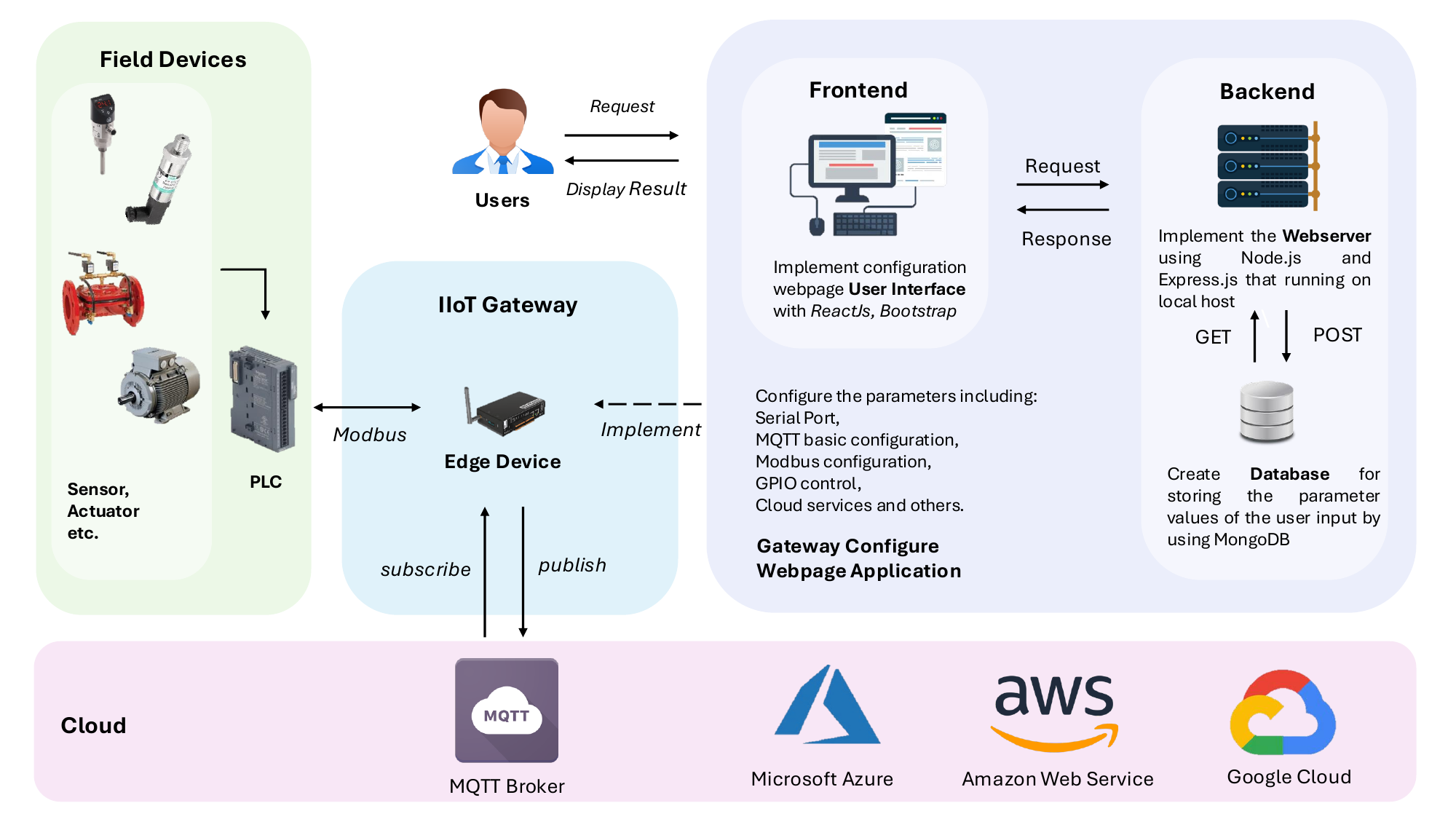}}
\caption{Comprehensive block diagram of the proposed modular software architecture, illustrating the data flow from physical Modbus nodes through the Application Middleware to cloud-based MQTT brokers.}
\label{fig1}
\end{figure*}

While these studies validate the hardware capabilities of SBCs, they often rely on static configuration scripts embedded within the operating system. This approach lacks the dynamic flexibility required for rapid reconfiguration on-site, posing barriers for non-experts and increasing errors. This paper addresses these challenges by proposing a modular database-driven software architecture tailored for open-source IIoT gateways that completely abstracts the underlying code. Our framework encapsulates low-level hardware configuration into a GUI-driven web application, empowering operators to reconfigure the system entirely through a graphical interface. This architecture accelerates the initial deployment phase and establishes a standardized, scalable foundation capable of supporting advanced edge-cloud orchestration in industrial automation.

\section{Proposed System Architecture}
In this section, we present our proposed system architecture for IIoT edge gateways. Our software\footnote{\url{https://github.com/peiyu16/Gateway_Configurator}} and tutorial videos\footnote{\url{https://tinyurl.com/4m324c6w}} have made publicly available.

Fig.~\ref{fig1} illustrates the high-level operational architecture of the proposed IIoT edge gateway. We separate the system into three distinct layers: (i) Edge Hardware Interface, (ii) Application Middleware, and (iii) Cloud Orchestration Layer. In this system, industrial field devices connect to the gateway via established fieldbus protocols, while the aggregated and processed data is transmitted to the cloud using MQTT.

.

\subsection{Hardware Infrastructure}
We utilized the IRIV PiControl system~\cite{cytron_iriv} as the primary edge node. This industrial-grade controller integrates a Raspberry Pi Compute Module 4 (CM4). It features isolated RS-485 interfaces, mitigating the risk of electrical noise disrupting serial communications in harsh factory environments. The hardware includes dual Gigabit Ethernet ports and expandable NVMe storage, providing the necessary bandwidth and memory for localized data buffering as shown in Table~\ref{tab1}.

\begin{table}[htbp]
\caption{Edge Gateway Hardware Specifications}
\begin{center}
\begin{tabular}{ll}
\toprule
\textbf{Component} & \textbf{Specification} \\
\midrule
Processing Unit & Quad-core Cortex-A72 (ARM v8) 64-bit @ 1.5GHz \\
Memory & 4GB LPDDR4-3200 SDRAM \\
Serial Interfaces & 2x Isolated RS-485, 1x RS-232 \\
Network & Gigabit Ethernet, Wi-Fi 802.11b/g/n/ac \\
Storage & M.2 NVMe SSD Support \\
\bottomrule
\end{tabular}
\label{tab1}
\end{center}
\end{table} 

\subsection{Application Middleware and Technology Stack}
The Application Middleware resides entirely on the CM4 and handles all computational logic, protocol translation, and state management. We developed the backend orchestrator using Node.js and the Express.js framework. Node.js utilizes an asynchronous, event-driven runtime environment, making it highly suitable for handling concurrent I/O operations without thread blocking. This efficiency is critical when simultaneously polling multiple RS-485 devices and maintaining a persistent WebSocket connection to the cloud.

We implemented MongoDB as the localized database engine. MongoDB's document-oriented NoSQL schema perfectly accommodates the hierarchical and often unstructured nature of IoT device configurations. The database stores network parameters, mapped Modbus registers, and user-defined automation rules. Storing this state locally ensures the edge gateway maintains autonomous operation even during severe network partitions, an industrial requirement for reliability.

We built the frontend interface utilizing React.js. This component-based JavaScript framework allows technicians to interact dynamically with the backend API. Users can assign I/O roles, define MQTT topics, and adjust scanning intervals through a responsive web dashboard, completely bypassing the Linux command line.

\subsection{Protocol Translation Workflow}
The software operates as a deterministic and bidirectional translation engine. During the uplink cycle, the Node.js polling engine iterates through the MongoDB collection of registered field devices. It dispatches read requests over the RS-485 bus utilizing the Modbus RTU protocol. Upon receiving the hexadecimal serial payload, the backend decodes the data based on user-defined function codes (e.g., Read Holding Registers). The system wraps the decoded values into a structured JSON payload and publishes it to the designated MQTT broker under specific Quality of Service (QoS) parameters.

During the downlink cycle, the Node.js client subscribes to command topics on the MQTT broker. When a cloud application issues an execution command, the middleware parses the JSON structure, translates it into the corresponding Modbus write command (e.g., Force Single Coil), and transmits it to the physical actuator.

\subsection{Automated Linkage Event Engine}
A localized linkage event engine is engineered to facilitate rapid, autonomous decision-making at the edge. Sending raw sensor data to the cloud for simple threshold evaluation introduces unnecessary latency and consumes bandwidth. 

We formalized the event engine as a finite state evaluation process. Let $S_t$ represent the continuous state of an analog sensor at time $t$. Let $\tau_{upper}$ and $\tau_{lower}$ represent the user-defined upper and lower execution thresholds. The system determines the discrete actuation state $A(S_t)$ of a coupled digital output device using the following piecewise function:

$$A(S_t) = \begin{cases} 1 & \text{if } S_t \geq \tau_{upper} \\ 0 & \text{if } S_t \leq \tau_{lower} \\ A_{prev} & \text{otherwise} \end{cases}$$

The Node.js backend evaluates this logic locally during every polling cycle. When $S_t$ breaches a threshold, the system triggers the corresponding digital output via the CM4 GPIO pins instantaneously, ensuring critical safety or operational thresholds are met regardless of cloud latency.

\section{Implementation and Results}
Our architecture is investigated through a series of structured integration tests simulating an industrial deployment. The preliminary outcomes are verified and reflected as follows.

\subsection{Testbed Configuration}
\begin{figure}[htbp]
\centerline{\includegraphics[width=0.9\columnwidth]{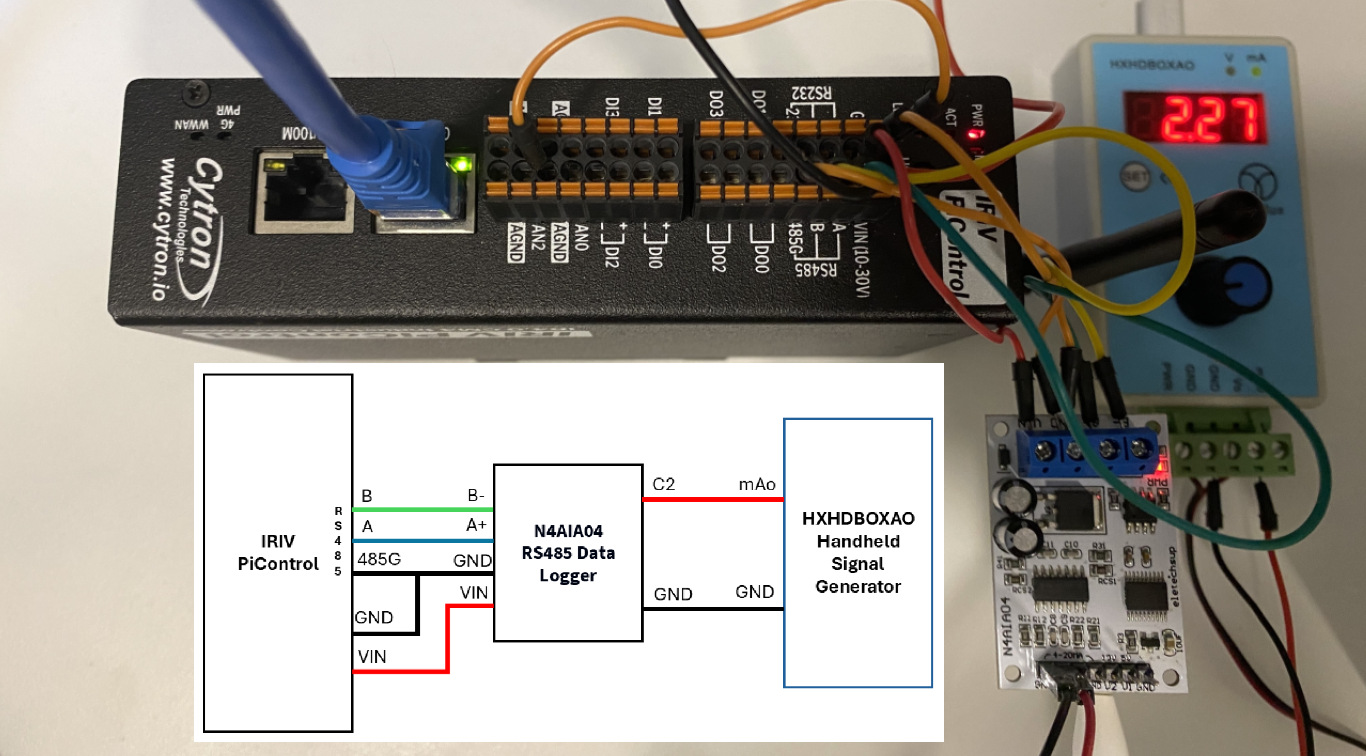}}
\caption{Physical testbed integration displaying the IRIV PiControl connected to the RS-485 Data Logger and Analog Signal Generator.}
\label{fig2}
\end{figure}

We constructed a physical simulation environment connecting a Handheld Signal Generator Adjustable Current Voltage Analog Simulator to an N4AIA04 RS-485 Data Logger as presented in Fig.~\ref{fig2}. This combination acted as a legacy Modbus field device continuously broadcasting fluctuating analog variables. The data logger is connected to the IRIV PiControl via the isolated RS-485 terminals. 

\begin{figure}[htbp]
\centerline{\includegraphics[width=0.95\columnwidth]{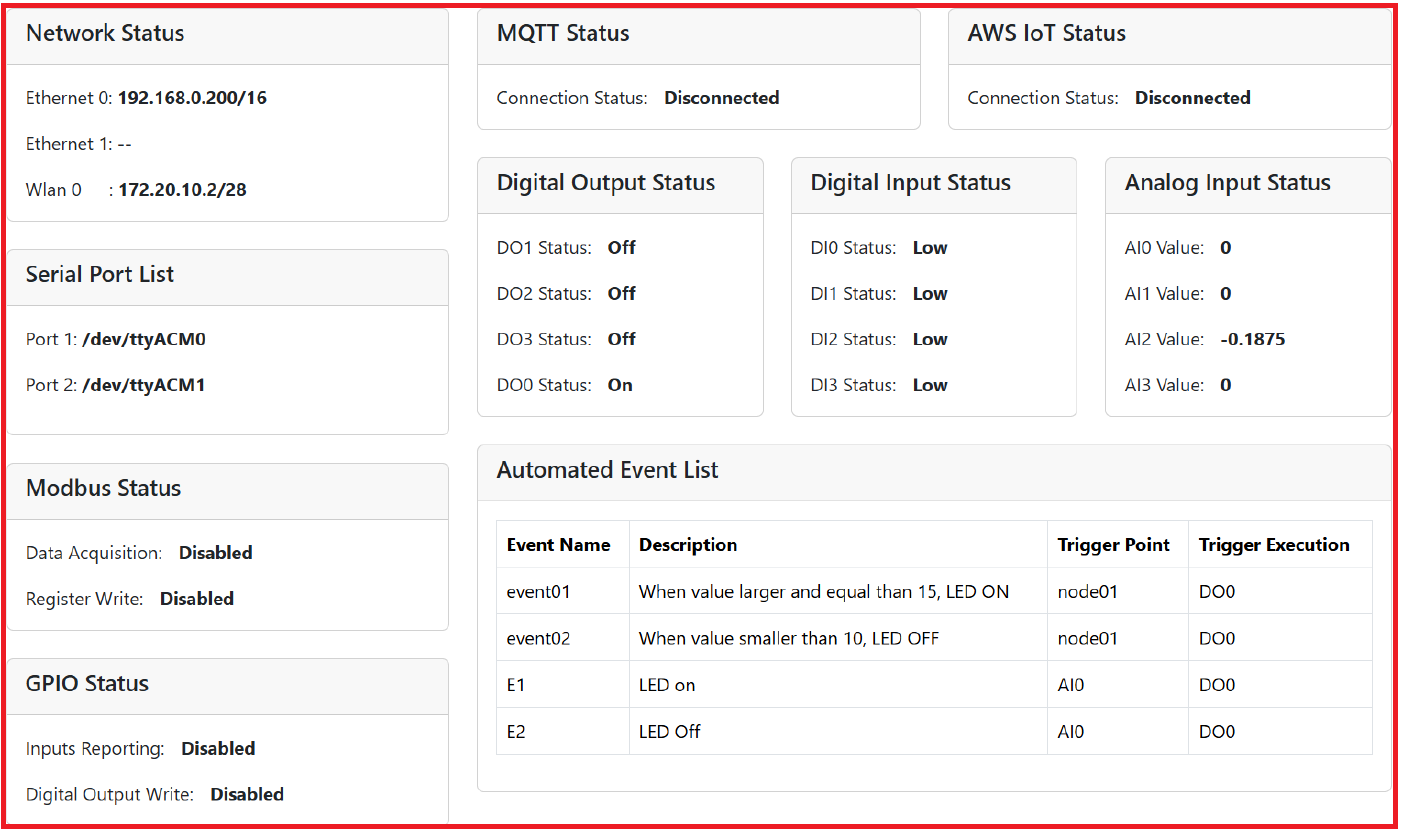}}
\centerline{\includegraphics[width=0.962\columnwidth]{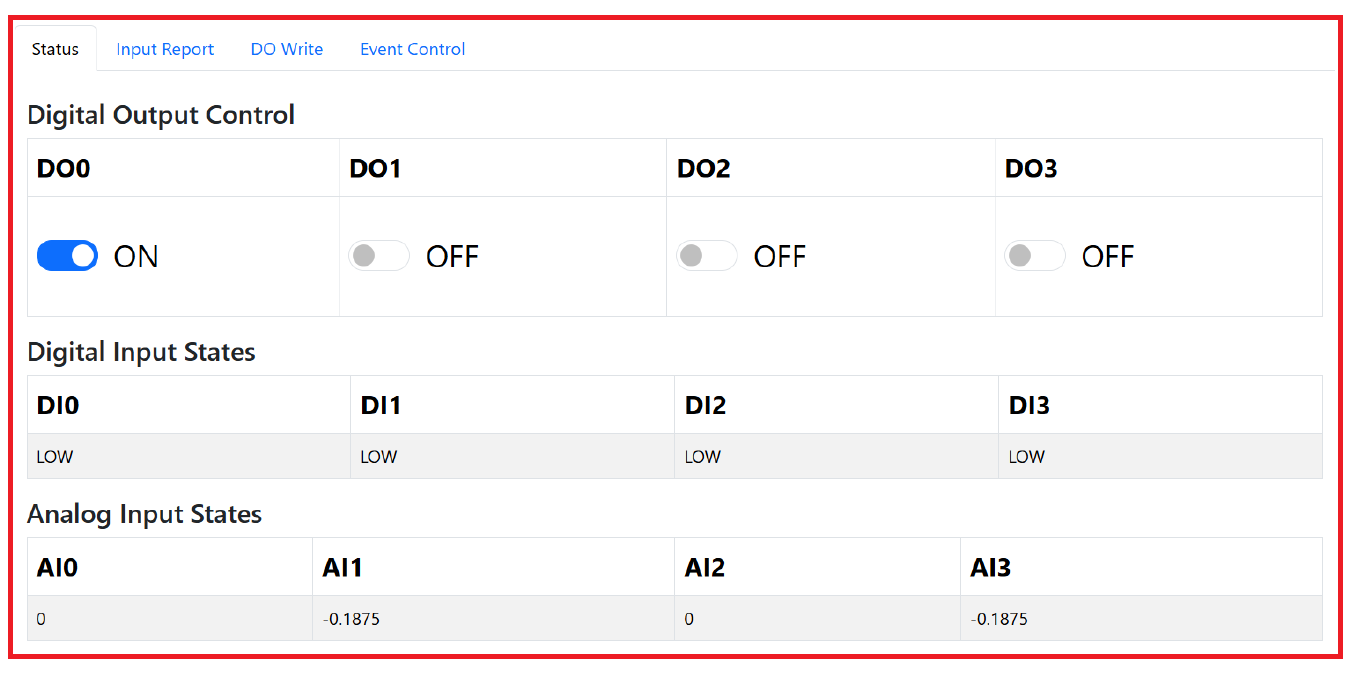}}
\centerline{\includegraphics[width=0.957\columnwidth]{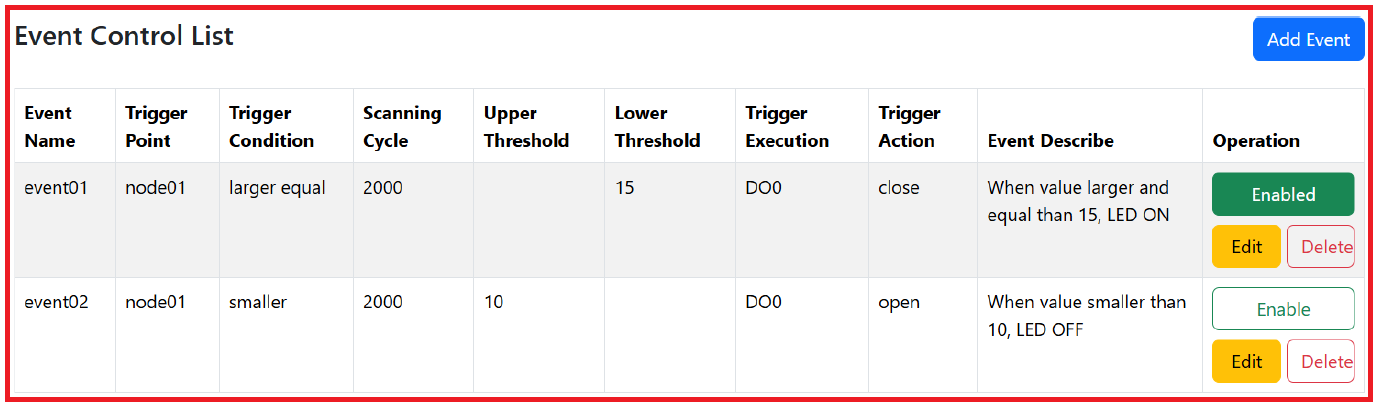}}
\caption{Web application: Dashboard, I/O status monitoring and event logs.}
\label{fig3}
\end{figure}

We also configured the gateway's React.js dashboard using an external laptop connected via Ethernet. The configuration process included defining the static IP address, mapping the data logger's holding registers, and establishing a TLS 1.2-secured connection to an AWS IoT Core endpoint.

\subsection{Performance and Usability Evaluation}
The software architecture demonstrated exceptional stability during the validation phase. The backend maintained a continuous polling cycle across the Modbus network while simultaneously publishing the JSON payloads to AWS IoT without packet loss. We also observed that the asynchronous Node.js backend maintained a minimal resource footprint. During continuous bidirectional communication, the CPU load on the ARM Cortex-A72 processor remained low, leaving substantial computational headroom for future integration.

The primary pilot achievement is the reduction in deployment overhead. Traditional command-line configuration of network interfaces, database schemas, and MQTT daemons on a Linux SBC typically requires several hours of expert engineering time. Utilizing our unified RESTful API and graphical interface, technicians can configure the entire data pipeline, from bare-metal hardware connection to active cloud telemetry, directly in the web GUI as shown in Fig~\ref{fig3}.

\section{Conclusions and Future Work}
In this paper, we presented a highly adaptable event-driven software architecture that simplifies the deployment of open-source IIoT edge gateways towards new network-centric paradigms such as Cloud-Fog Automation. This framework addresses the complexities of manual coding by abstracting serial protocol management and network configuration into a responsive web interface. The system effectively bridges the gap between legacy industrial equipment and modern cloud infrastructure. The integration of localized threshold evaluation further ensures deterministic response times for critical operations, hence reducing reliance on cloud processing.

Despite demonstrating reliable performance in device management and automated event controls, the current system scope has several limitations. The hardware dependency on RS-485 for physical connectivity restricts network scalability in remote or highly distributed deployments. Additionally, the software currently translates exclusively between Modbus and MQTT, limiting interoperability with other prevalent industrial communication standards.

Future work will address these limitations by expanding both the hardware connectivity and the software's computational capabilities. We will integrate configuration support for extended wireless modules such as LoRa and 4G/5G, to leverage the full connectivity suite and NVMe storage capacity of the edge controller. We will implement security mechanisms for the edge-based containerized framework~\cite{mahmud2025trusted} and focus on the horizontal scalability of this software framework by incorporating agentic automation protocols in era of Large AI models~\cite{jiang2026comprehensive, nguyen2025task, boateng2025survey}. This advancement will enable dynamic task placement and collaborative decision-making across a distributed swarm of networked edge gateways, which enables SMBs and industrial practitioners to seamlessly transition into next-generation AI-driven ICPS.

\bibliographystyle{IEEEtran}
\bibliography{refs}

\end{document}